\documentclass[12pt,prd,showpacs,psfig,dcolumn,epsfig,superscriptaddress]{revtex4}
\usepackage{amsmath}
\usepackage{amssymb}
\usepackage{graphicx}
\usepackage{dcolumn}% Align table columns on decimal point
\usepackage{bm}% bold math
\usepackage{amsfonts}

\makeatletter
\@ifundefined{textcolor}{}
{%
 \definecolor{BLACK}{gray}{0}
 \definecolor{WHITE}{gray}{1}
 \definecolor{RED}{rgb}{1,0,0}
 \definecolor{GREEN}{rgb}{0,1,0}
 \definecolor{BLUE}{rgb}{0,0,1}
 \definecolor{CYAN}{cmyk}{1,0,0,0}
 \definecolor{MAGENTA}{cmyk}{0,1,0,0}
 \definecolor{YELLOW}{cmyk}{0,0,1,0}
}

\makeatother

\begin{document}
\title{Search for doubly charmed hadron at B factory}

\author{Yi Jin}
\affiliation{School of Physics and Technology, University of Jinan, Jinan 250022,
P. R. China}
\author{Shi-Yuan Li}
\affiliation{School of Physics, Shandong University, Jinan 250100, P. R. China}
\author{Yan-Rui Liu}
\affiliation{School of Physics, Shandong University, Jinan 250100, P. R. China}
\author{Zong-Guo Si}
\affiliation{School of Physics, Shandong University, Jinan 250100, P. R. China}
\author{Tao Yao}
\affiliation{School of Physics, Shandong University, Jinan 250100, P. R. China}

\begin{abstract}
The doubly charmed hadron production at B factories is of special importance for the study of the hadron structure and the color connections before hadronization. To suppress the combination background fluctuations of the reconstructed hadron mass spectra, we suggest a three-jet event shape trigger. After these three jets are identified by their energy and angular distributions, it is found that: 1) The background process $e^+ e^- \to c\bar {c} \to h's$ in consideration of the final hadron system $\Lambda_c^+K^-\pi^+ +X$ are significantly suppressed. 2) For the selected events, about half of the particles, $\Lambda_c^+$, $K^-$, $\pi^+$, which obviously can not belong to the decay products of doubly charmed hadron, can be vetoed.
The relevant hadronization is investigated.
\end{abstract}
\pacs{12.38.Bx, 13.87.Fh, 24.10.Lx}
%\textbf{Keywords:} {color Connection, Doubly Heavy Baryon, Three-jet Event, String Effect}\\

\date{\today}\maketitle
%%%%%%%%%%%%%%%%%%%%%%%%%%%%%%%%
\section{Introduction}\label{sec1}
%%%%%%%%%%%%%%%%%%%%%%%%%%%%%%%%
In the hadroproduction process of charged hyperon beam on nuclear targets, a resonance decaying into $\Lambda_c^+K^-\pi^+$ \cite{SELEX} and $pD^+K^-$ \cite{SELEX2} is observed by the SELEX Collaboration. The resonance could be one of doubly charmed hadrons, $\Xi_{cc}^+$ baryon.
%The SELEX Collaboration claimed the observation of $\Xi_{cc}^+$ decay into $\Lambda_c^+K^-\pi^+$ \cite{SELEX} and $pD^-K^-$ \cite{SELEX2} in the hadroproduction of charged hyperon beam on nuclear targets.
However, in the following studies of the $\Lambda_c^+K^-\pi^+$ mass spectrum at B factories, both Barbar \cite{BABAR} and Belle \cite{BELLE} have yet not found this resonance. A search performed by FOCUS Collaboration in the photoproduction process gives no evidence, either \cite{FOCUS}. Recently, LHCb Collaboration look for $\Xi_{cc}^+$ in the decay channel $\Lambda_c^+K^-\pi^+$ in $pp$ collisions, and find no signal \cite{LHCb}. In an improved search with more data and additional decay modes, Belle Collaboration again fail to give positive evidence \cite{BELLE2}. The production mechanism for doubly charmed baryons seems intriguing.

All the above mentioned experiments have used the $\Lambda_c^+K^-\pi^+$ channel to search for $\Xi_{cc}^+$. The number of reconstructed $\Lambda_c^+$ in BaBar ($\sim 600000$) and that in FOCUS ($\sim  19500$) are both much larger than that in SELEX ($\sim 1650$), but only SELEX observes doubly charmed baryon. This fact forces people to suspect whether $\Xi_{cc}^+$ can be produced only in hadroproduction with hyperon beams. It is necessary to perform more exploration at Belle and BaBar to dispel the suspicion. In $e^+e^-$ annihilation, the cross sections do not prefer the forward direction as in high energy hadronic interactions. The 4$\pi$ spectrometers can record most of the reaction processes. If we improve the method to veto the background and/or fluctuations for the measurement, we are possible to get a signal or a stronger exclusion.

In searching for $\Xi_{cc}^+$ in the $\Lambda_c^+K^-\pi^+$ channel (in this paper we take $\Xi_{cc}^+$ as the example, and the case for the corresponding antiparticles is exactly the same), the tracking procedure from final hadrons is as follows.
First, $\Lambda_c^+$ is reconstructed from the decay channel $pK^-\pi^+$. Then $\Xi_{cc}^+$ is searched for via the invariant mass spectrum
of the $\Lambda_c^+K^-\pi^+$ system in which the $K^-$ and $\pi^+$ are selected between the collision point and $\Lambda_c^+$ decay vertex.
This approach is the general one in finding new hadrons, but it may be inefficient in searching for doubly charmed hadrons at $e^+e^-$ colliders.
One explicit observation is that all produced particles and their strong decay products can be between the two vertices mentioned above. If no specified restriction on the phase space, the cross section of $e^+e^- \to c \bar c \to h's$ is much larger than that of $e^+e^- \to c \bar c c \bar c$ by orders of magnitude. So most of $\Lambda_c^+$'s are produced in the background process rather than $\Xi_{cc}^+$ decay.
 There are also $K^-$'s and $\pi^+$'s belonging to the `$X$' but not from $\Xi_{cc}^+$ decay located between the collision point and the $\Lambda_c^+$ decay vertex, even for $e^+ e^- \to \Xi_{cc}^+ + X$ production process. They can introduce fatal combination background fluctuations to mass spectrum.
  Therefore, the key point is to veto as much as possible `wrong' $\Lambda_c^+$ as well as $K^-$'s and $\pi^+$'s from combinations, so that one can get the clean invariant mass spectrum.

\begin{figure}[htb]
\centering
\scalebox{0.35}{\includegraphics{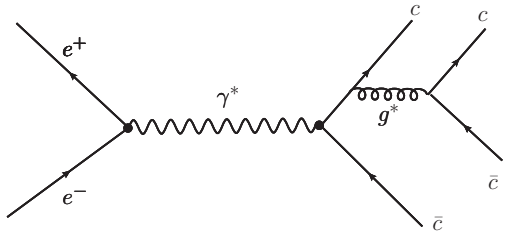}}
\caption{One of the eight lowest order Feynman diagrams of the process $e^+e^- \to c\bar{c} c\bar{c}$, the other diagrams can be obtained by changing the vertex of gluon to the antiquark line and exchanging the momenta between two $c$'s (two $\bar {c}$'s).
%(b)One of the four lowest order Feynman diagrams of the process $e^+ e^- \to c\bar {c} q\bar{q}$,  the other diagrams can be obtained by changing the vertex of gluon to the antiquark line and $c \leftrightarrow q$ ($\bar c \leftrightarrow \bar q$) exchange.
%%Hereafter, $q$ represents a light quark.
}\label{ccqq1}
\end{figure}

\begin{figure}[htb]
\scalebox{0.9}{\includegraphics{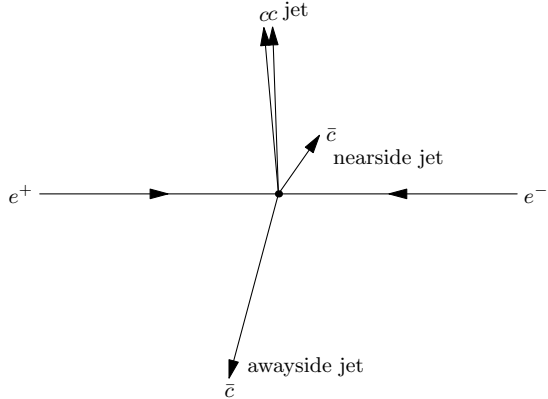}}
\caption{A typical schematic phase structure in the three-jet shape for the partonic level of the process $e^+e^- \to c\bar{c} c\bar{c} \to \Xi_{cc}^+ + X$ in the center of mass frame, of which the two $c$'s are close to each other in phase space and they can hadronize into $\Xi_{cc}^+$. 
%Here we require that the momenta of the two $c$'s are almost the same.
%A schematic diagram for $\Xi_{cc}^+$ production in the center of mass frame.
}\label{schematic}
\end{figure}

In $e^+e^-$ annihilation, one must have two pairs of $c\bar {c}$ to produce $\Xi_{cc}^+$. The lowest order partonic process is $e^+e^- \to c\bar{c} c\bar{c}$ with the Feynman diagrams shown in FIG.~\ref{ccqq1}. One $c\bar{c}$ pair is produced via the virtual photon, and the other $c\bar{c}$ pair is then produced via the virtual gluon splitting which is emitted from one of the quark/antiquark lines.
When these two $c$'s are close to each other in phase space, they can hadronize into $\Xi_{cc}^+$. Otherwise, all the charm quarks/antiquarks will hadronize into singly charmed hadrons. The latter case also belongs to the background. The quark and the gluon propagators in FIG.\ref{ccqq1} together determine the phase space configuration. A typical schematic phase structure for the process $e^+e^- \to (cc) \bar{c} \bar{c}$ of the former case in the center of mass frame is shown in FIG.\ref{schematic}.
%It is almost determined by the propagators of the quark and the gluon.
This apparent feature of the Feynman diagrams suggests that the generation of $\Xi_{cc}^+$  associates with the three-jet like event shape: a $cc$ jet, one $\bar{c}$ jet close to it, while the other $\bar{c}$ jet almost in the opposite direction. Therefore, one may experimentally identify such three jets first, then reconstruct $\Lambda_c^+$ only in the $cc$ jet (and the nearside $\bar{c}$ jet if not well separated), and then search for doubly charmed baryons employing these $\Lambda_c^+$, $K^-$'s and $\pi^+$'s only in the same jet. By such an investigation, those $\Lambda_c^+$'s as well as $K^-$'s and $\pi^+$'s which apparently belong to the awayside $\bar{c}$ jet could be vetoed. 
 The essential question is how to identify these three jets. In this work we find that in most cases the $cc$ jet is the most energetic one, while the awayside $\bar{c}$ jet is the second. Further more, these three jets belong to three energy regions almost without overlap, which means that all these three jets can be well identified. All the particles (like those $\Lambda_c^+$'s, $K^-$'s and $\pi^+$'s) belong to the awayside $\bar{c}$ jet can be vetoed from the construction of $\Lambda_c^+$ as well as $\Xi_{cc}^+$. 
 This method can also suppress the background process by orders of magnitude.

As usual, one can realize the above results at hadronic level. However, the conventional event generators have never considered the hadronization of this special case with a special color connection.
For the four-charm-quark system $c\bar{c}c\bar{c}$, in $e^+e^-$ annihilation, if two $c$'s belong to different color-singlet clusters, doubly charmed baryon can not be produced. So as is pointed out in \cite{Ma:2003zk,xicc} that the production of such a baryon is a fingerprint of a special kind of  color connection. We will investigate the color connections and model the hadronization.

This paper is organized as follows. Sec. \ref{sec2} is devoted to the calculations on the jet distributions at the partonic level. In Sec. \ref{sec3}, we investigate the corresponding hadronization for the related doubly charmed hadrons production in $e^+ e^-$ annihilation. Finally, we provide a short summary in Sec. \ref{sec4}.

\section{Jet distributions at partonic level}\label{sec2}

Jet definition is infrared safe, and the perturbative calculation can coincide with the hadronic one.
Here we take the Jade algorithm \cite{JADE}
\begin{equation}
y_{ij}=\frac{(p_i+p_j)^2}{E_{cm}^{2}}\label{cutoff-1}
\end{equation}
to define the jets. The parameter $y_{cut}$ is thus introduced and two partons/particles are considered as being in one jet when $y_{ij}<y_{cut}$. In the following calculations, we apply this jet algorithm to $e^+e^- \to c\bar{c} c\bar{c}$ process by the following equation:
%and the background system of $c\bar {c} q\bar{q}$ respectively to obtain the corresponding three-jet cross section $\sigma_{3-jet}$:
 \begin{equation}
 d{\sigma}^{c\bar{c} c\bar{c}}_{3-jet}=d\sigma[\Theta(M_{c}+M_{c}+\delta m-M_{cc})+\Theta(M_{\bar{c}}+M_{\bar{c}}+\delta m-M_{\bar{c}\bar{c}})].
 \end{equation}
% \begin{equation}
% d\tilde{\sigma}^{c\bar {c} q\bar{q}}_{3-jet}=d\sigma_b[\Theta(M_{c}+M_{q}+\delta m-M_{cq})+\Theta(M_{\bar{c}}+M_{\bar{q}}+\delta m-M_{\bar{c}\bar{q}})].
% \end{equation}
 In the above equation, $d\sigma$ is calculated by employing the Feynman diagrams in FIG.\ref{ccqq1}.
 $\Theta (x)$ is the Heaviside step function. $\delta m$ is an important parameter which gives the value range of the invariant mass $M_{cc}$ ($M_{\bar{c}\bar{c}}$). Its concrete value will affect the prediction for the production rate of $\Xi_{cc}^+$. However, the value of $\delta m$, whenever it is small comparing to $\sqrt{y_{cut}}E_{cm}$, will not affect the event shape. Here we take $\delta m=1.0$ GeV, as an example.

In our calculations, we take the fine structure constant to be
$\alpha=1/137$, the strong coupling constant to be $\alpha_{s}$ = 0.12 and the mass of charm quark to be $m_{c}$ = 1.5 GeV/$c^{2}$. We give two sets of results corresponding to two $y_{cut}$ values: $y_{cut1}=0.08$ and $y_{cut2}=0.2$. The former is very small since $\frac{(2M_c)^2}{E_{cm}^{2}}\sim 0.08$; while the latter is fairly large. The background process $e^+ e^- \to c\bar {c} \to h's$ is calculated by PYTHIA \cite{pythia} with default parameters.

\begin{figure}[htb]
\centering
\begin{tabular}{cccc}
\scalebox{0.18}{\includegraphics{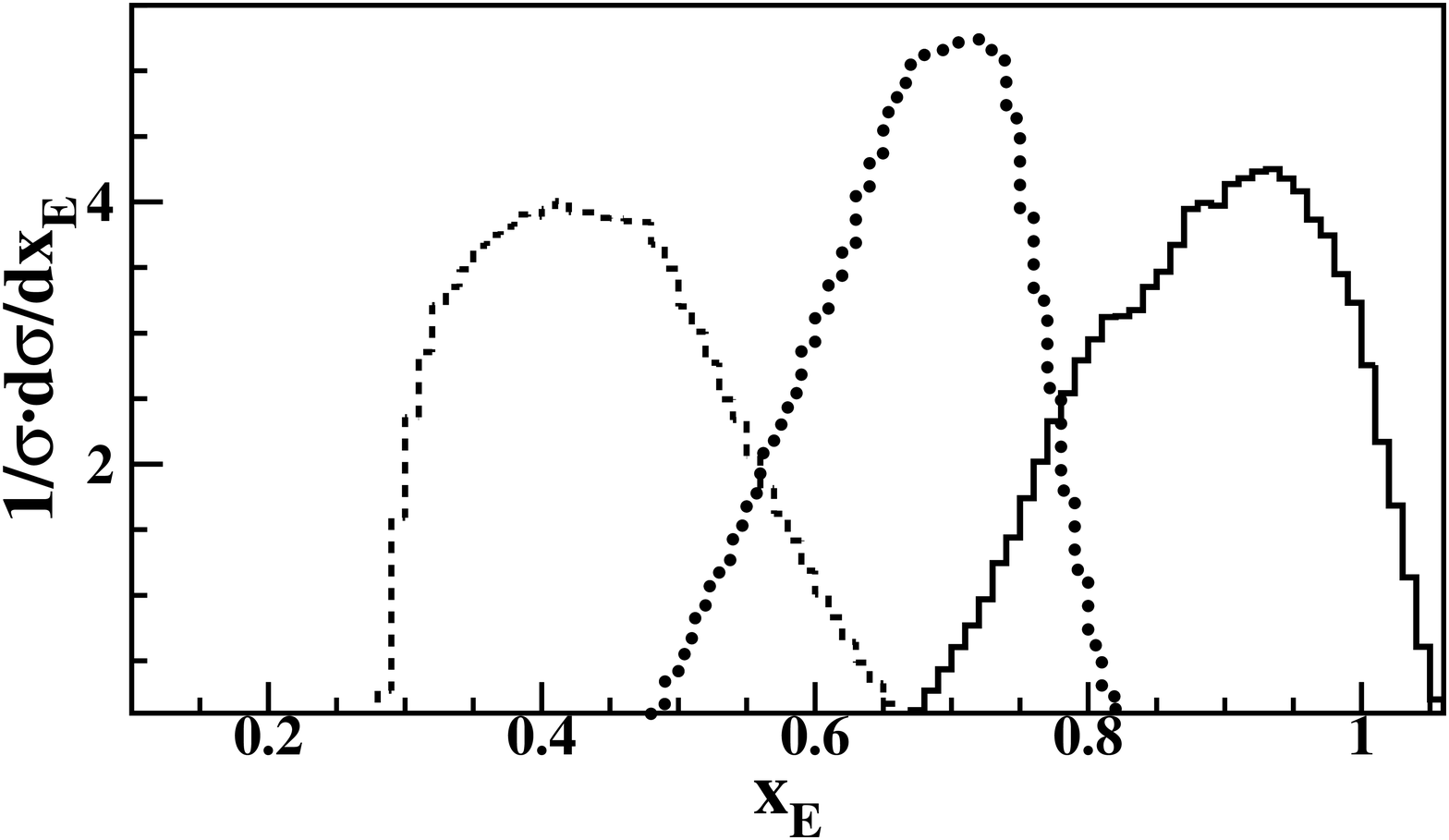}}&
\scalebox{0.18}{\includegraphics{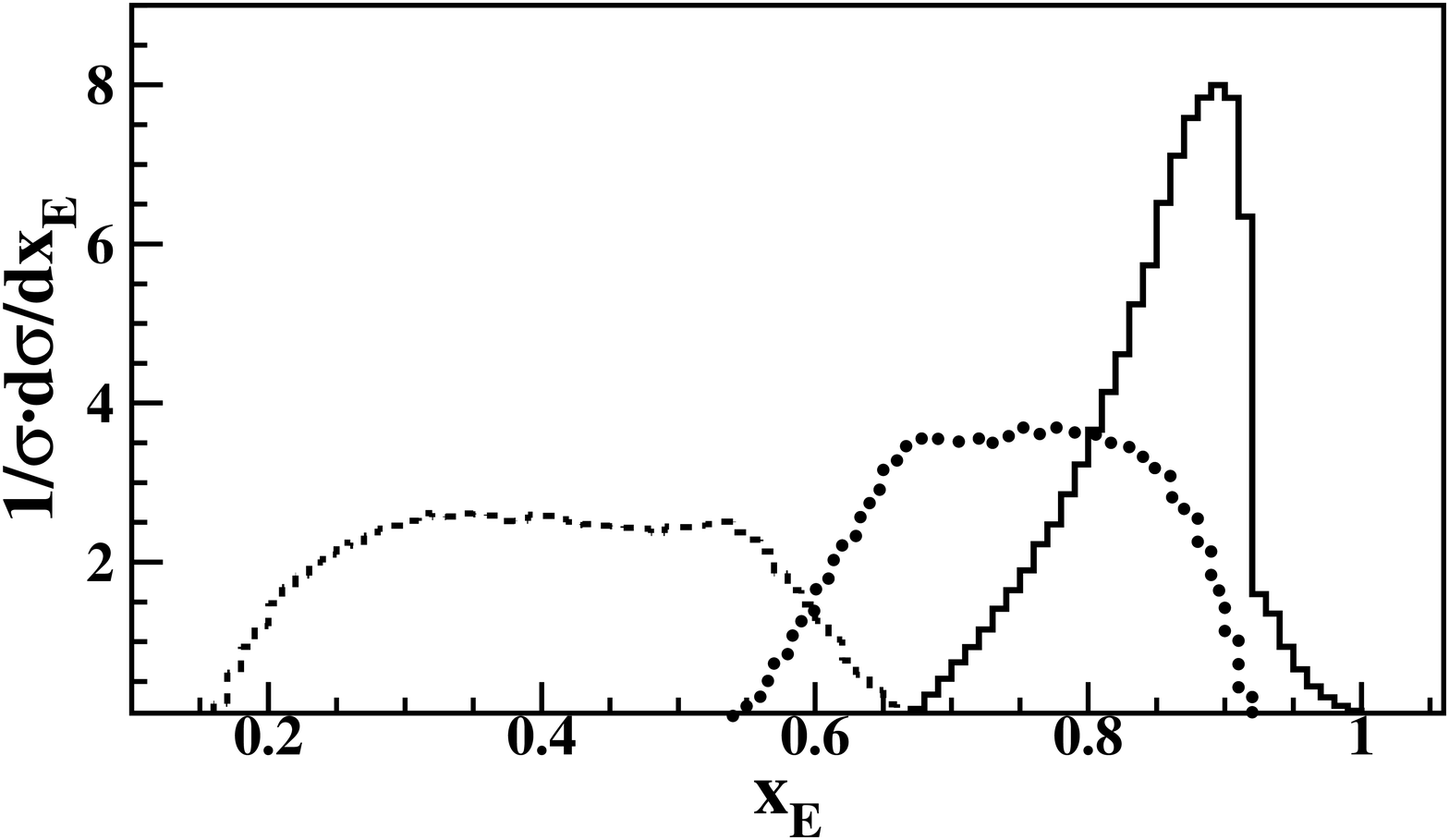}}&\\
{\scriptsize (a)}&{\scriptsize (b)}\\
\scalebox{0.18}{\includegraphics{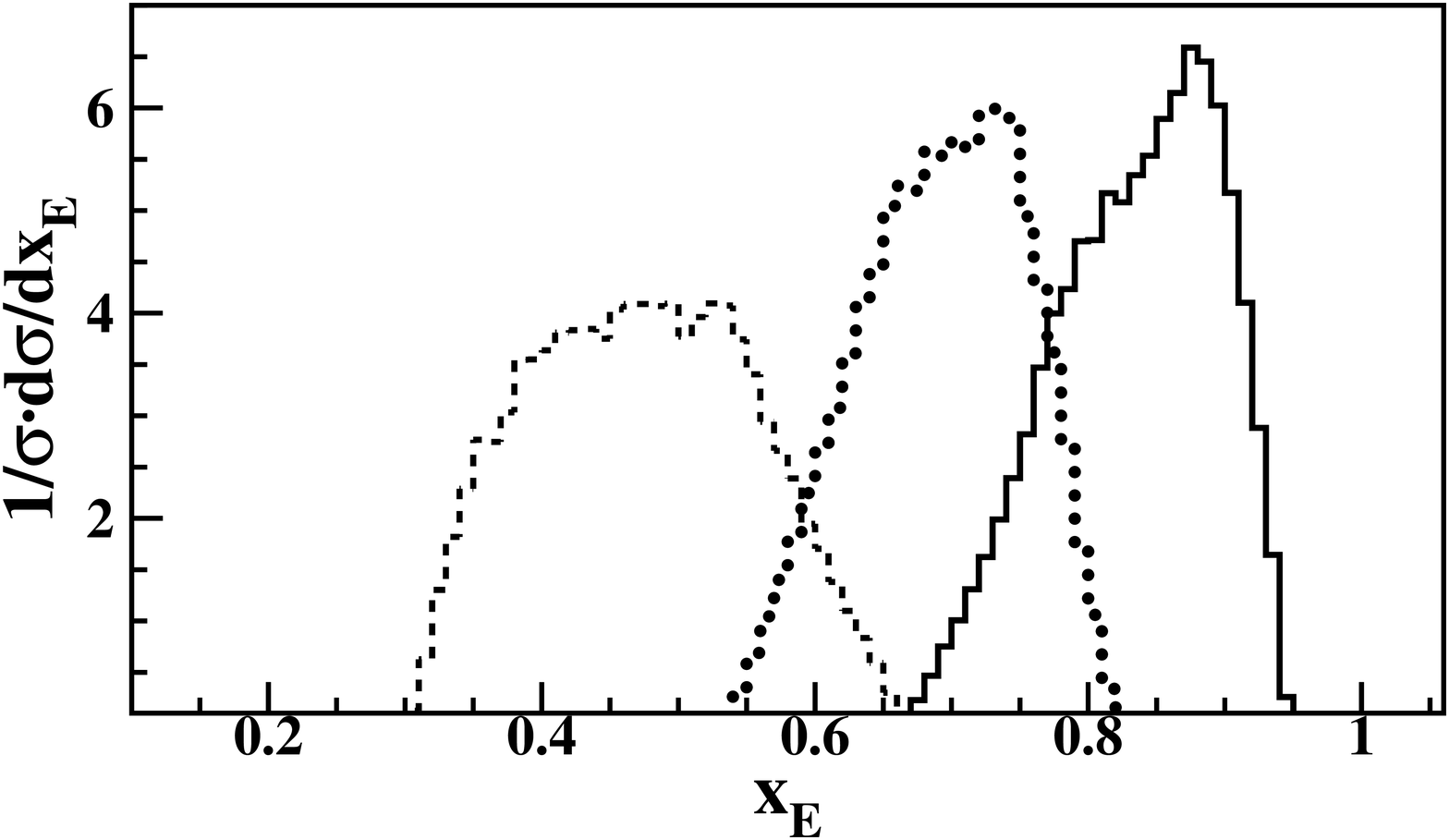}}&
\scalebox{0.18}{\includegraphics{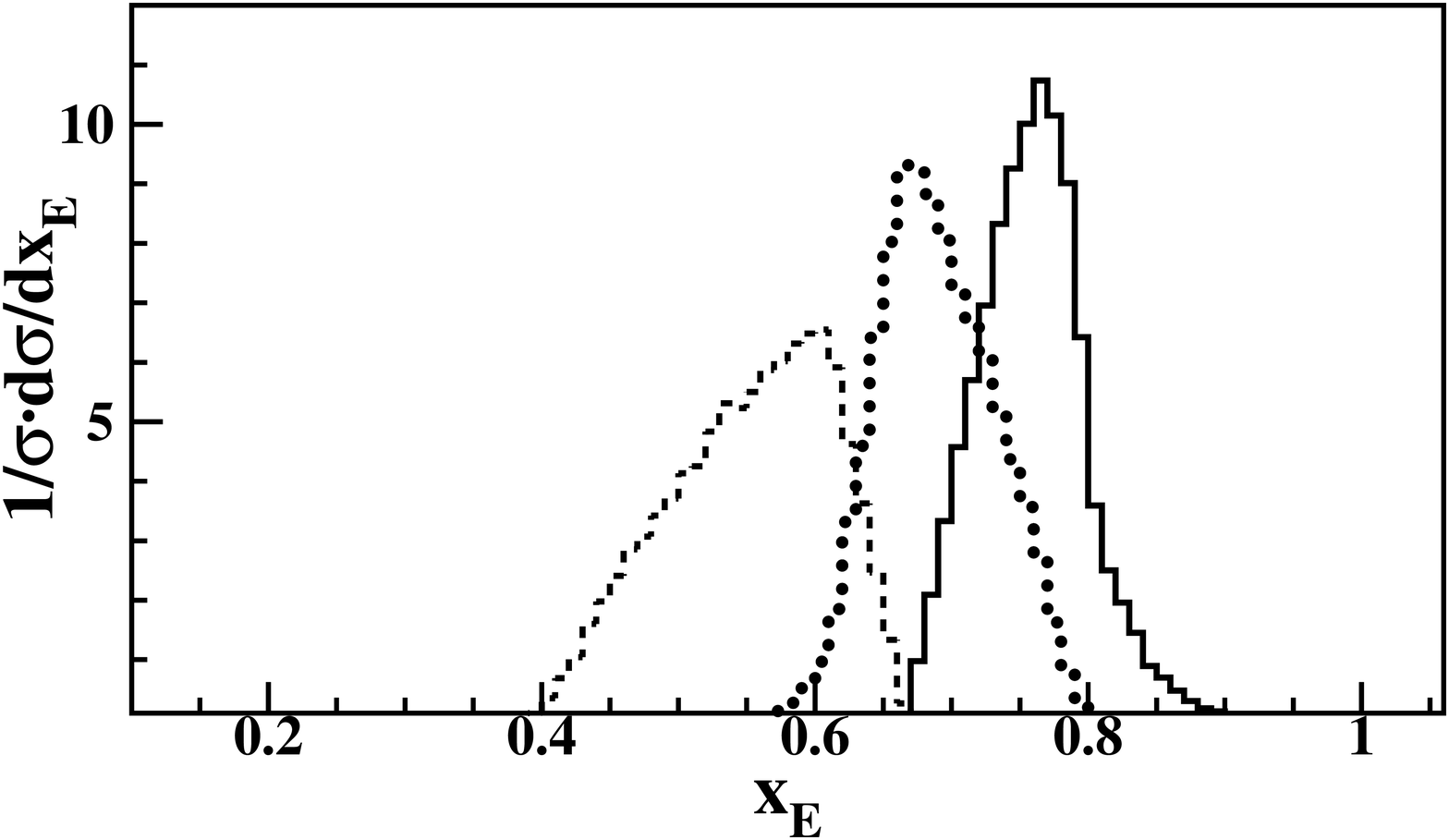}}\\
{\scriptsize (c)}&{\scriptsize (d)}
\end{tabular}
\caption{
The energy fraction distributions for jet-1/2/3 (solid/dotted/dashed line). Here we employ the scaled dimensionless variable $x_{E}=2E/\sqrt{s}$. (a) $c\bar{c} c\bar{c}$ with $y_{cut1}$=0.08; (b) background process with $y_{cut1}$=0.08; (c) $c\bar{c} c\bar{c}$ with $y_{cut2}$=0.2; (d) background process with $y_{cut2}$=0.2.}
\label{xe008lmh}
\end{figure}

FIG. \ref{xe008lmh}(a) and (c) show the energy fraction distributions for the three jets. The energies of these three jets are well ordered, and the energy ranges have almost no overlap. Hereafter, we will adopt `jet-1/2/3' to represent the most/second/least energetic jet.
So by such an ordering in energies, we can identify each jet. To improve the purity, we will set the requirement that the energy fractions %$x_{E}$
of the jet-1, 2 and 3 fall into (0.80, 1.05), (0.55, 0.80) and (0.30, 0.55) respectively in the same time. Without these constraints, the percentage of the events that jet-1 just contains the $cc$ is 86.7 (80.2) for $y_{cut1}$=0.08 ($y_{cut2}$=0.2). After the energy range constraints are considered, this percentage increases to 98.2 (97.2). By this way ({\it i.e.}, the energy ordering-range constraint), we can keep almost all the events needed for the reconstruction for the $\Xi_{cc}^+$.
The most attracting advantage is that the $K^-$'s and $\pi^+$'s as well as $\Lambda_c^+$, which belong to the second energetic jet-2 and obviously are not from $\Xi_{cc}^+$ decay, can be straightforwardly vetoed.

\begin{figure}[htb]
\centering
\scalebox{0.3}{\includegraphics{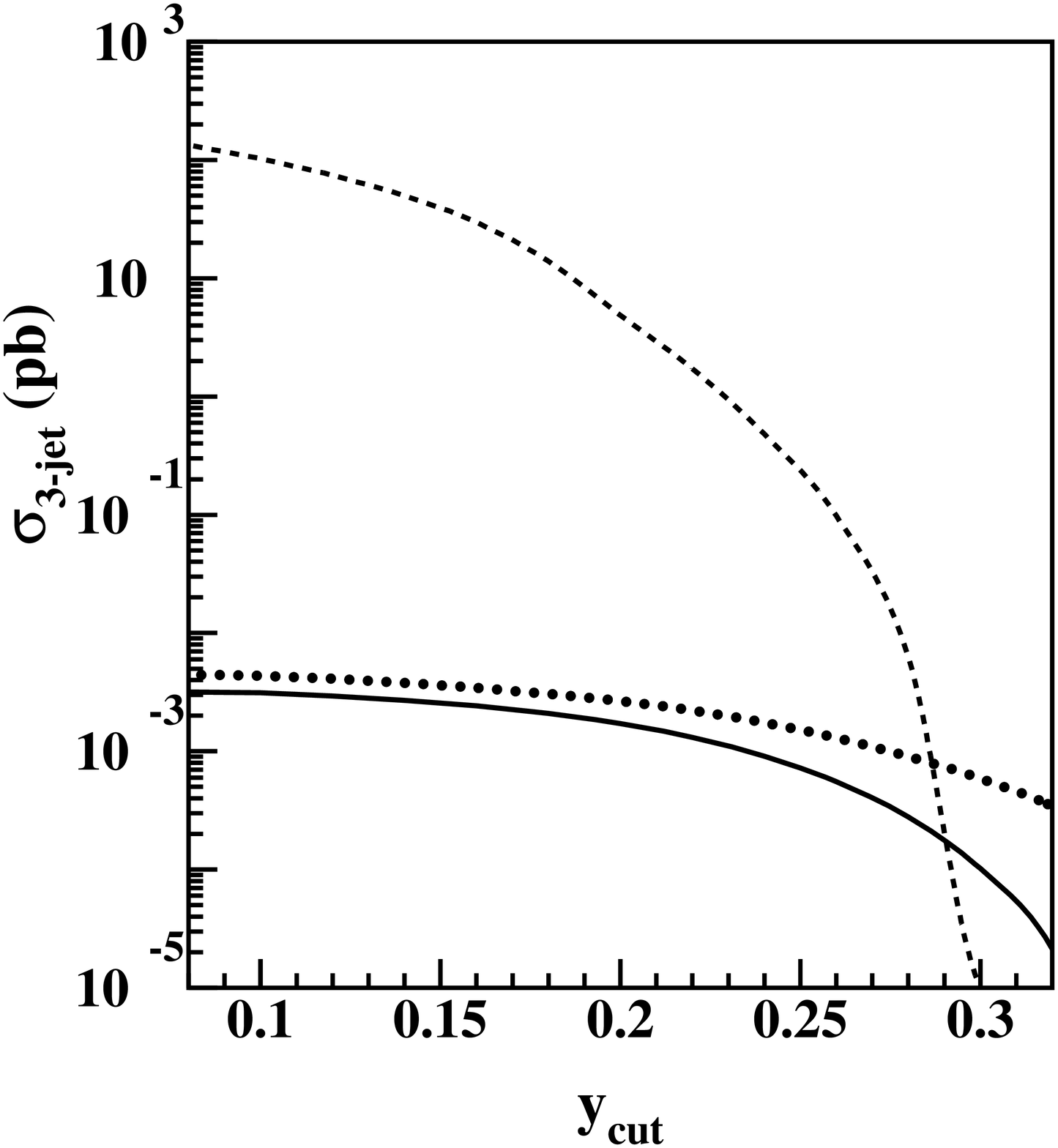}}
\caption{The three-jet cross section $\sigma_{3-jet}$ with respect to $y_{cut}$ for $\delta m=1.0$ GeV. The solid/dotted line is for the signal process of $e^+ e^- \to c\bar{c} c\bar{c}$ with/without the energy ordering-range constraint, and the dashed line is for the background process of $e^+ e^- \to c\bar {c} \to h's$ with the energy ordering-range constraint.}
\label{psigmajd}
\end{figure}

The above results for $e^+e^- \to c\bar{c} c\bar{c}$ are insensitive to the $y_{cut}$. The three-jet cross section $\sigma^{c\bar {c} c\bar{c}}_{3-jet}$ of $e^+ e^- \to c\bar {c} c\bar{c}$ process varies from $4.45 \times 10^{-3}$ pb to $2.66 \times 10^{-3}$ pb for $y_{cut}$ from 0.08 to 0.2. Moreover, after the energy ordering-range constraint is employed, $\sigma^{c\bar {c} c\bar{c}}_{3-jet}$ decreases slightly and varies from $3.17 \times 10^{-3}$ pb to $1.71 \times 10^{-3}$ pb for $y_{cut}$ from 0.08 to 0.2. 
While the three-jet cross section of the background process $e^+ e^- \to c\bar {c} \to h's$ is sensitive to $y_{cut}$ as well as the energy ordering-range constraint. This can be seen from FIG. \ref{xe008lmh} and \ref{psigmajd}.
For a small $y_{cut1}=0.08$ with the energy ordering-range constraint, the background cross section is about 134 pb, which decreases to 4.9 pb for $y_{cut1}=0.2$. With the design integrated luminosity 50 ab$^{-1}$ of SuperKEKB, the corresponding signal/noise ratio varies from 1.9 to 5.5. This means that the number of the events needed to be analyzed decreases much for the large $y_{cut}$. 

 %So even important is that, less background data lead to less fluctuation of the mass spectrum.
 
% For the large $y_{cut2}=0.2$, itself is very efficient to constrain the background $\sigma^{c\bar {c} q\bar{q}}_{3-jet}$ to $2.05 \times 10^{-2}$ pb which decreases one order of magnitude compared to the one for $y_{cut1}=0.08$, and the energy ordering-range constraint further decreases it to $6.27 \times 10^{-3}$ pb. Now the event number of the background is comparable to that of the signal.
 
\begin{figure}[htb]
\centering
\scalebox{0.3}{\includegraphics{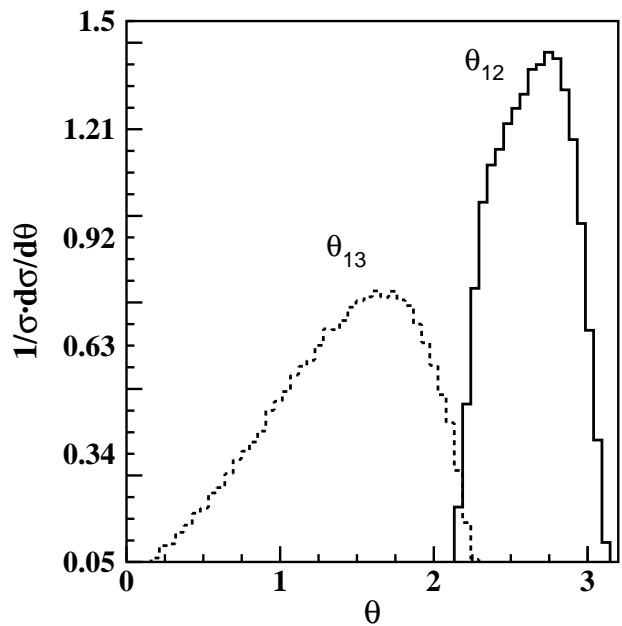}}
\caption{The distributions of the relative angle $\theta_{12}$ (solid line) which is between jet-1 and jet-2, and $\theta_{13}$ (dashed line) which is between jet-1 and jet-3 with $y_{cut}$=0.2.
The momentum of the jet-1 is chosen to be the $x$-axis. By defining jet-1,2,3, we have set the energy range constrain. }
\label{theta1213}
\end{figure}

 We know that to measure the invariant mass spectrum, more pairs of combination not only heavily enlarge the quantity of the work, but also introduce much background fluctuations which lead to the signal overlooked. As is well known, in the center of mass frame of $e^{+}e^{-}$ annihilation, the momenta of the three jets must be in the same plane due to momentum conservation \cite{Ellis:1976uc}.
We can easily investigate the distributions of the relative angle $\theta_{12}$ (between jet-1 and jet-2) and $\theta_{13}$ (between jet-1 and jet-3).
The corresponding results are displayed in FIG. \ref{theta1213}.
The angular distributions of jet-2 and jet-3 again do not overlap, and jet-2 is clearly the awayside jet.
Then we can further identify these jets and veto the background by requiring the angular ranges.
For example, jet-2 is well separated from jet-1, and jet-3. The distribution $\theta_{13}$ peaks around $\pi/2$, so one can further veto some contributions from jet-3 for the events with large $\theta_{13}$.

In the above three-jet events,  the created $cc$ pair with small invariant mass could hadronize into  $\Xi_{cc}^+$ but also other particles, {\it e.g.}, $T_{cc}$ if they exist. The quark content and quantum numbers of the ground $T_{cc}$ are $cc\bar{u}\bar{d}$ and $I(J^{P})=0(1^+)$, respectively. The difference between the productions of $\Xi_{cc}^+$ and $T_{cc}$ lies in the hadronization process.  All kinds of hadronization effects will be dealt with in the next section.

\section{color connections and hadronizations}\label{sec3}

As pointed in \cite{hlsy}, the color connection of the production process $e^+e^- \to c\bar{c} c\bar{c}$ has the following interesting property:
\begin{equation}
(3_{1}\otimes3_{2})\otimes(3_{1}^{*}\otimes3_{2}^{*})=(3_{12}^{*}\oplus6_{12})\otimes(3_{12}\oplus6_{12}^{*})=(3_{12}^{*}\otimes3_{12})\oplus(6_{12}\otimes6_{12}^{*})\oplus\cdots,\label{diqcl}
\end{equation}
where $6_{12}$ ($6_{12}^{*}$) denotes the sextet (anti-sextet) representation of the $SU_{c}(3)$ group. When two (anti)quarks in
color state $3^{*}(3)$ attract each other and form a(n) `(anti)diquark'
and their invariant mass is sufficiently small, such a cluster has a certain probability
of hadronizing like a(n) (anti)diquark, as discussed in our previous work \cite{hlsy},
by triggering the leading baryons and two-jet-like event shape. For the case in which only one of the pair has a small invariant mass, the color configuration can be better written as
\begin{eqnarray}
(3_{1}\otimes3_{2})\otimes3_{1}^{*}\otimes3_{2}^{*}=3_{12}^{*}\otimes3_{1}^{*}\otimes3_{2}^{*}\oplus\cdots, & or\nonumber \\
3_{1}\otimes3_{2}\otimes(3_{1}^{*}\otimes3_{2}^{*})=3_{1}\otimes3_{2}\otimes3_{12}\oplus\cdots.
\label{q2qq}
\end{eqnarray}
The above property should be taken into account in considering the hadronization model of the process $e^+e^- \to c\bar{c} c\bar{c} \to \Xi_{cc}^+ + X$.

In Eq. \ref{q2qq}, the color configuration as a whole is like a `big baryon.'
The $cc$($\bar{c}\bar{c}$) with a small invariant mass can hadronize into a(n) (anti)baryon (tetraquark) as a(n) (anti)diquark.
In this case, the diquark $cc$ must combine with a quark $q$ (antidiquark) to form a  color-singlet system, of which with a large invariant mass, the hadronization is a `branching' process via the creation of quarks from the vacuum by the strong interactions within the system.
The created quarks and the primary quarks are combined into color-singlet hadrons.
Here, we take the diquark $cc$ as an example. The $cc$ needs a quark/antidiquark to form $\Xi_{cc}^+$/$T_{cc}$. To balance the quantum numbers of color and flavor, an antiquark/diquark must be simultaneously created from the vacuum. To branch them further, more
quark pairs and diquark pairs must be created from the vacuum via
the interactions among the quark system. Such a cascade process
will proceed until the end of time, when most of the `inner energy'
of the entire system is transformed into the kinematical energies
and masses of the produced hadrons. Each of two newly created
quarks (antidiquarks) combines with each of the primary $\bar{c}$'s to respectively hadronize into two open charmed hadrons, which can be described by an assigned concrete hadronization model (for details, see \cite{string,pythia,cluster,herwig}). Because of the success of the Lund string model \cite{string}, especially its realization by PYTHIA/JETSET \cite{pythia}, the above hadronization procedure can be easily realized.
For the configurations considered here, the above process is
straightforward, except that for each step, we must assign special
quantum numbers for each specific kind of hadron according to its
production rate.

\begin{figure}[htb]
\centering
\scalebox{0.21}{\includegraphics{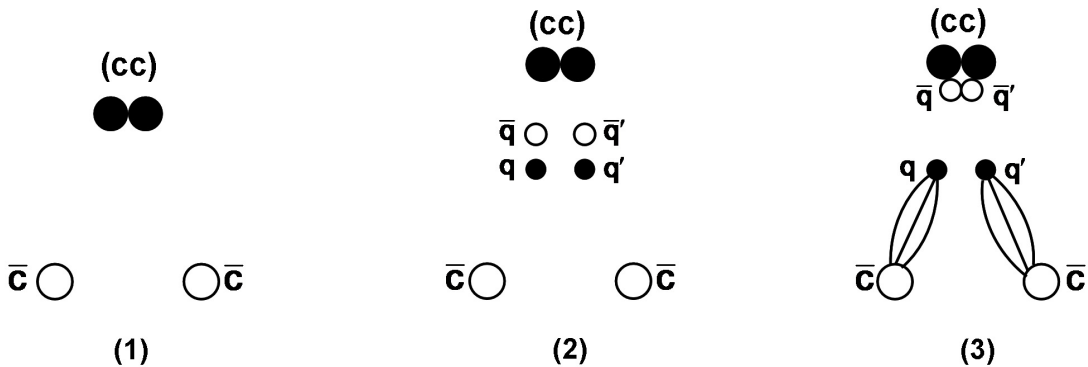}}
\caption{$T_{cc}$ production and the string formation for the hadronization of the $(cc)\bar{c}\bar{c}$ system with the aid of quark creation from the vacuum. Solid circles represent quarks, while hollow circles represent antiquarks.
The primary $\bar c \bar c$ connect to quarks respectively via two strings in (3).}
\label{cc2tcc}
\end{figure}

For the case of $T_{cc}$ production, if the complementary diquark pair is broken by the
interactions within the remaining system and then each becomes
connected to the other two primary $\bar{c}$'s to form
two strings, the resultant hadronization can be described by the
conventional string-fragmentation picture, see FIG.\ref{cc2tcc}.

\begin{figure}[htb]
\centering
\scalebox{0.35}{\includegraphics{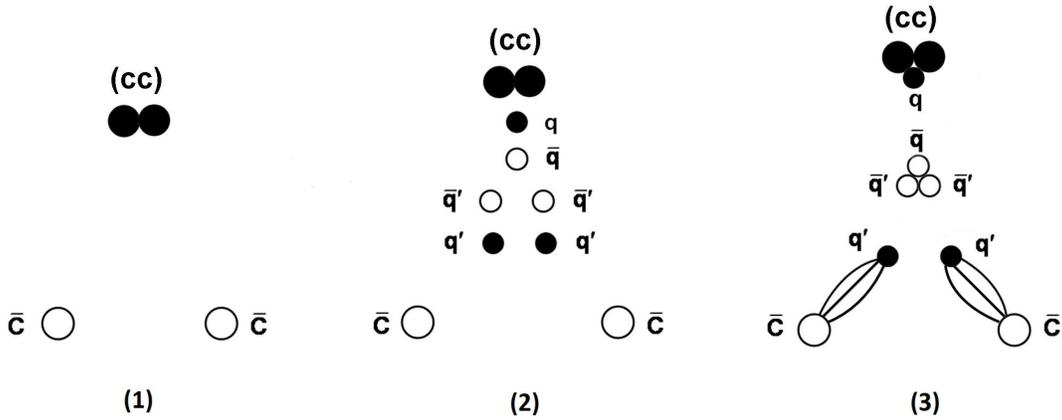}}
\caption{$\Xi_{cc}^+$ production and the string formation for the hadronization of the $(cc)\bar{c}\bar{c}$ system with the aid of quark creation from the vacuum.
The primary $\bar c \bar c$ connect to quarks respectively via two strings in (3).}
\label{cc2xi}
\end{figure}

For the case of $\Xi_{cc}^+$ production, the complementary antiquark can produce
an antibaryon by combining with an antidiquark, and then the
balancing diquark can help to form two strings in the same manner
described above. This procedure is illustrated in FIG.\ref{cc2xi}.

The fragmentation of the heavy diquark can be described by the Peterson formula \cite{pet83}
\begin{equation}
f(z)\propto\frac{1}{z(1-1/z-\epsilon_{Q}/(1-z))^{2}},\label{Peterson}
\end{equation}
where $\epsilon_{Q}$ is a free parameter, which is expected to scale between
flavors as $\epsilon_{Q}\propto1/m_{Q}^{2}$. In the following, we will show the results corresponding to $\epsilon_{Q}=1/25$ for the $cc$ diquark.

To describe the fragmentation of the complementary (anti)
quark, we adopt the fragmentation function employed by the
LUND group \cite{and83a}
\begin{equation}
f(z)\propto z^{-1}(1-z)^{a}exp(-bm_{\perp}^{2}/z),\label{Anderson}
\end{equation}
where a and b are free parameters. In our program, we take $a=0.3$ GeV$^{-2}$
and $b=0.58$ GeV$^{-2}$, as used in PYTHIA \cite{pythia}. Another
topic that we do not discuss here in detail is the excited states of doubly charmed hadrons. If we consider the fact that heavy quark
masses fatally break the SU(4) and/or SU(5) flavor symmetries and
we assume that all the excited states dominantly decay into the
ground state, the details of the differences can be neglected.
In the following, we give the numerical results for $\Xi_{cc}^+$ and $T_{cc}$.
The fragmentation of the strings can be referred to the classical book
about the Lund model by B. Anderson \cite{string} and the PYTHIA manual \cite{pythia}.

\begin{figure}[htb]
\centering
\scalebox{0.3}{\includegraphics{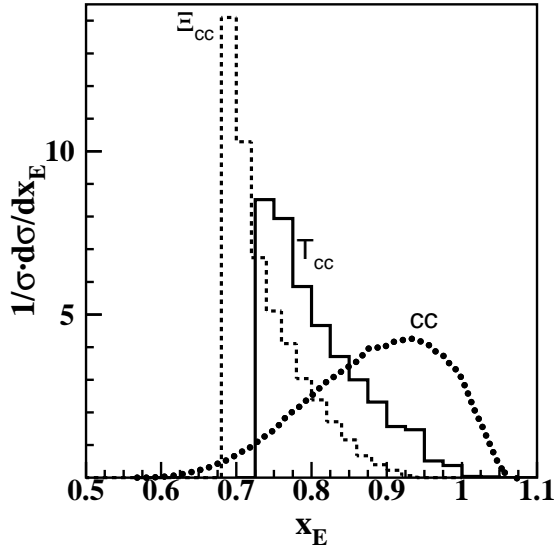}}
\caption{The distribution of the energy fraction for $\Xi_{cc}^+$ and $T_{cc}$ compared to the diquark $cc$ as a function of the scaled dimensionless variable $x_{E}=2E/\sqrt{s}$. The smooth dotted line represents the diquark $cc$, and the dashed/solid line represents $\Xi_{cc}^+$/$T_{cc}$.}
\label{xe}
\end{figure}

We just show the energy distributions of $\Xi_{cc}^+$ and $T_{cc}$ in FIG. \ref{xe} to demonstrate the hadronization effects. The absolute values of the distributions more or less depend on the parameters in Eqs.(\ref{Peterson},\ref{Anderson}).
The fragmentation functions and/or the parameters can be tuned by comparison to data once they are obtained.
As seen from the energy-fraction distributions, one of the most significant hadronization effects can be observed in the fragmenting process. $\Xi_{cc}^+$ and $T_{cc}$ only take part
of the energy of the $cc$ according to Eq.(\ref{Peterson}).

We would like to emphasize that the above description of the formation of these two strings is only one of the effective descriptions of the branching procedure of the quark system. In principle, this and other effective models can yield the same event-shape at the hadronic level.

With all these analysis, it is clear that the method to veto `wrong' particles and eliminate background process  hence is  applicable to the searches for other doubly heavy hadrons, {\it e.g}., $T_{cc}$ without doubt. %The quark content and quantum numbers of the ground $T_{cc}$ are $cc\bar{u}\bar{d}$ and $I(J^{P})=0(1^+)$, respectively.
The existence of $T_{cc}$ has been investigated in various approaches \cite{Zouzou,Lipkin,Heler,Brac,Stancu,Bielich,Vijande,Ebert,SHLee,Zhang,Yang}. The obtained mass is around the $DD^*$ threshold, depending on models. The excited $T_{cc}$ state with $cc$ in the sextet color representation is also expected \cite{Tccexotic}.
The production of doubly charmed tetraquarks in various colliders has been discussed in Refs. \cite{Ma:2003zk,xicc,Tccexotic,Tccprod,ExHIC}. If the predicted mass of $T_{cc}$ is above the $DD^*$ threshold, it can decay strongly into $DD^*$ and one may search for it in this channel. If it is below the $DD^*$ threshold, a detectable weak decay channel is $D^+K^-\pi^+$. Now Belle experiment is searching for $T_{cc}$ states. If Belle gives no evidence of them, a search in the weak decay channel with the above mentioned method is proposed. The situation of non-observation might change if one can veto the background contributions by identifying those three jets firstly and searching for $T_{cc}$ mainly in the $cc$ jet.

\section{Summary}\label{sec4}
In $e^+e^-$ annihilation, the event shape of doubly charmed hadron production is dominantly a three-jet one. By this event shape trigger, the background can be suppressed by orders of magnitude. Furthermore, our results clearly demonstrate that the jet which contains the doubly charmed baryon is the most energetic one, with its energy in the range almost without overlap with the other two. So it is easy to be identified. The mediate energetic jet can be well separated by the energy range as well as the angle relative to the first one. So all the particles belong to it can be vetoed from the reconstruction of the invariant mass. Thus the efficiency will be significantly increased, while the combination fluctuations will be significantly decreased in searching for $\Xi_{cc}^+$/$T_{cc}$.
%This method is not sensitive to the choice of the $y_{cut}$ and the hadronization process.
We have set up  the hadronization model, so more of the final hadron distributions or effects can be studied further and retuned once we have the data showing the signal exist.

\section*{Acknowledgments}

This work is supported in part by Natural Science Foundation of China and Shandong Province, SRF for ROCS, SEM. We thank Cheng-Ping Shen for the discussion on $T_{cc}$ study at Belle.

%This work is supported by NSFC(11047029, 11275114, 11275115), SRF for ROCS, SEM, and Natural Science Foundation of Shandong Province. The authors thank all of the members in Theoretical Particle Physics Group of Shandong University for their helpful discussions.

\end{document}